\documentclass[a4paper,amsmath,amssymb,preprintnumbers]{jpconf}
\usepackage{graphicx}
\usepackage{subfigure}

\def\beq{\begin{equation}}
\def\eeq{\end{equation}}
\def\bea{\begin{eqnarray}} 
\def\eea{\end{eqnarray}}

\def\mev{{\rm MeV}}

\def\eps{\varepsilon}

\begin{document}
\title{Parity Violation by a Dark Gauge Boson}
\author{Hye-Sung Lee}
\address{Theory Division, CERN, CH-1211 Geneva 23, Switzerland}
\ead{hyesung.lee@cern.ch}

\begin{abstract}
We overview the dark parity violation, which means the parity violation induced by a dark gauge boson of very small mass and coupling.
When a dark gauge boson has an axial coupling, as in dark $Z$ model, it can change the effective Weinberg angle in the low-energy experiments such as the atomic parity violation and the low-$Q^2$ polarized electron scatterings.
Such low-energy parity tests are an excellent probe of the dark force.
\end{abstract}

\section{Introduction}
In this article\footnote{This article and the presentation at PAVI 14 meeting (Skaneateles, New York, July 14-18, 2014) are mainly based on the works and discussions with H. Davoudiasl and W. Marciano at Brookhaven National Lab, USA.}, we overview the dark parity violation \cite{Davoudiasl:2012ag,Davoudiasl:2012qa,Davoudiasl:2014kua}, which means the parity violation induced by a dark gauge boson.
The dark gauge boon (we use $Z'$ for its notation) is a hypothetical particle with a very small mass and a small coupling to the Standard Model (SM) particles.
There are multiple motivations for the dark gauge interaction (or dark force) including the dark matter related ones, but we will focus only on the $g_\mu - 2$ motivation here.
(See Ref.~\cite{Essig:2013lka} for a recent review on the subject.)
We will first discuss the relation of the $g_\mu - 2$ and the dark gauge interaction, and briefly overview the typical dark force searches in the labs.
Then we will discuss the dark force searches through the low-energy parity test.

\section{Muon anomalous magnetic moment and dark gauge interaction}
\begin{figure*}[tb]
\begin{center}
\subfigure[]{
      \includegraphics[width=9pc,clip]{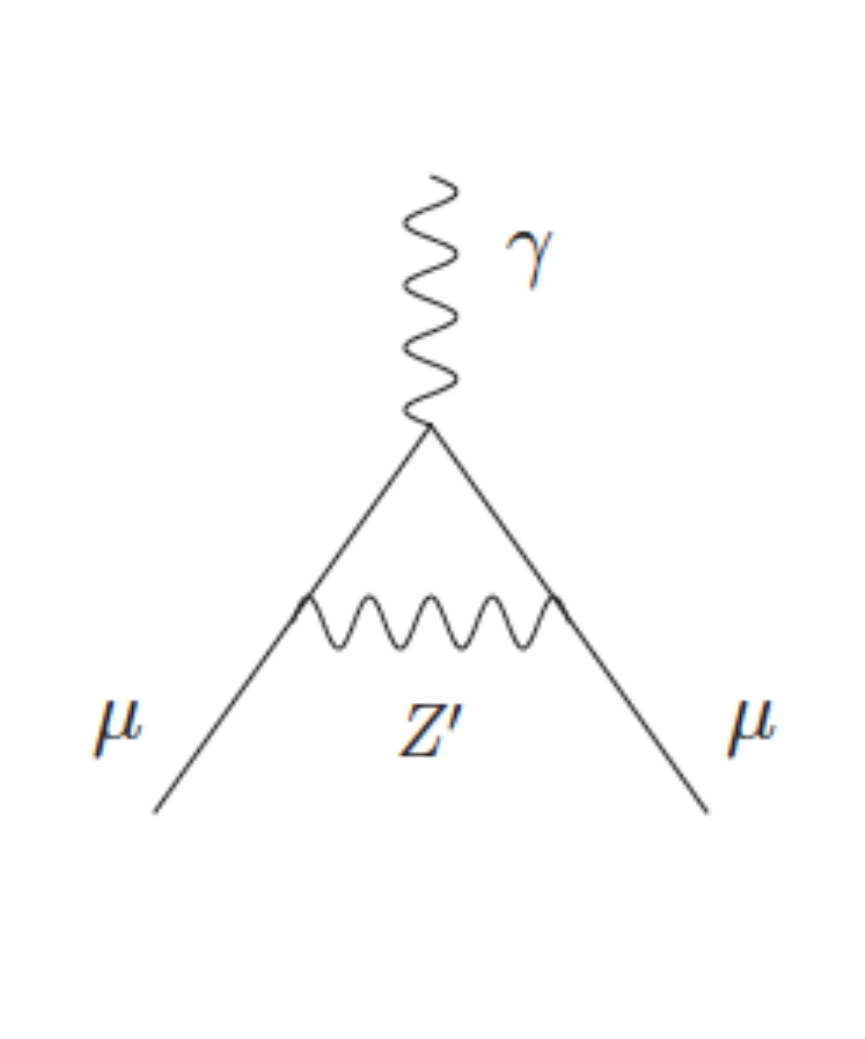}
      \label{fig:g-2A}
} \quad
\subfigure[]{
      \includegraphics[width=15pc,clip]{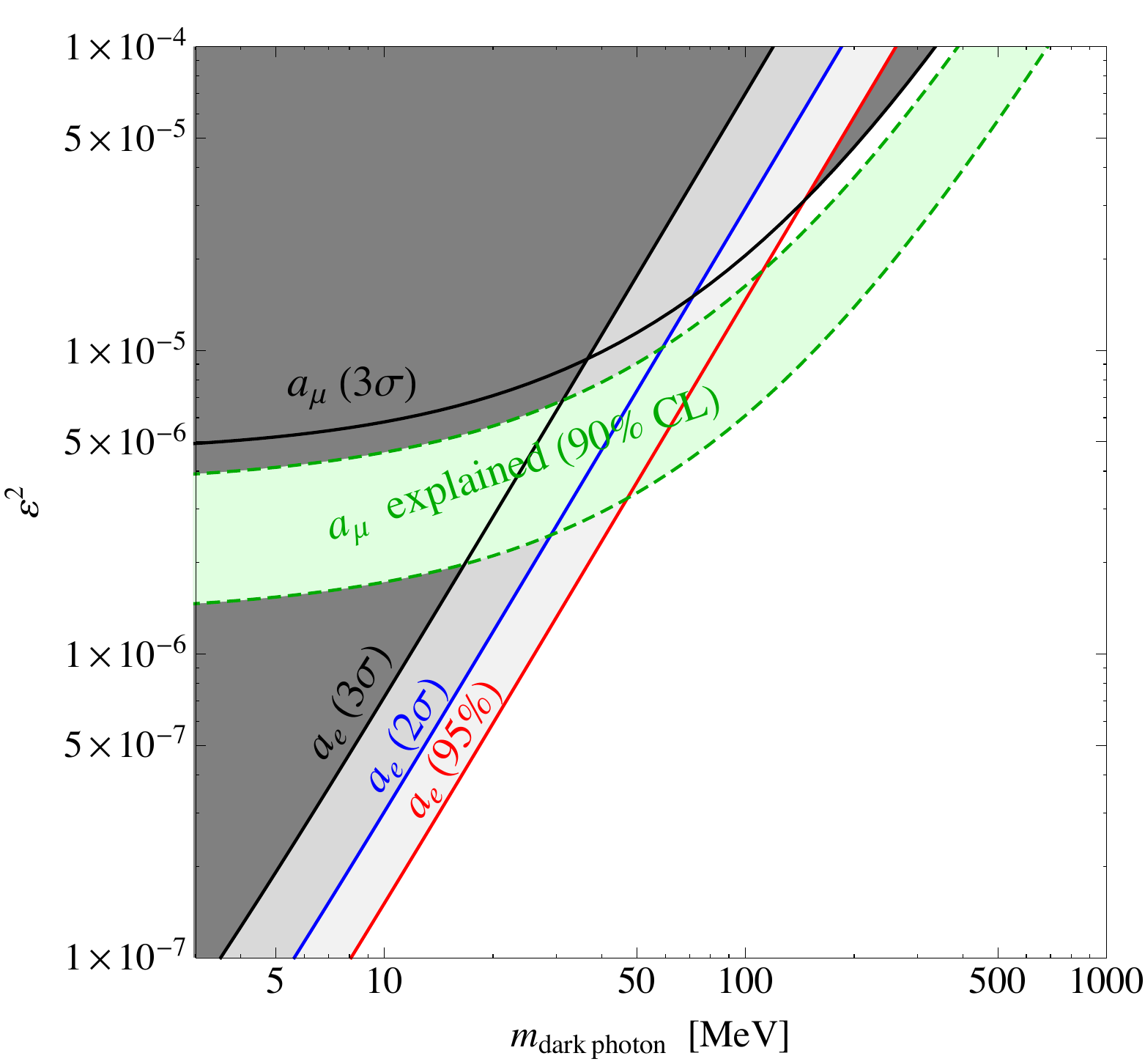}
      \label{fig:g-2B}
}
\caption{(a) Dark gauge boson contribution to the $a_\mu$. (b) Dark photon parameter space with $a_e$ and $a_\mu$ constraints. The green band is a parameter region motivated by the $3.6 \sigma$ C.L. anomaly in the $a_\mu$. The effective coupling constant of the dark photon is $\alpha' = \eps^2 \alpha_{EM}$.}
\end{center}
\end{figure*}

The $g_\mu - 2$ issue has been always an important motivation and the constraint for the new physics.
Currently, there is a $3.6 \sigma$ C.L. discrepancy between the measurement and the SM prediction in the muon anomalous magnetic moment, $a_\mu \equiv (g_\mu - 2)$ \cite{Agashe:2014kda,Bennett:2006fi}.
It is one of the major motivations for a new light gauge boson too as it can contribute to the $g_\mu - 2$ through a loop-correction [Fig.~\ref{fig:g-2A}].
Unlike the most other motivations, it is independent of the unknown dark matter properties, and it is also independent of the $Z'$ decay branching ratios.
Figure \ref{fig:g-2B} (from Ref.~\cite{Davoudiasl:2014kua}) shows the parameter space of the dark photon along with the $g_\mu - 2$ favored parameter space (green band) as well as the constraints from $a_e$ and $a_\mu$.

To explain the $a_\mu$ deviation, which is a positive quantity, the $Z'$ coupling to the muons should be vector-coupling dominated because it would give the right sign for the $Z'$ contribution to the $a_\mu$ to explain the discrepancy, while the axial coupling would give the wrong sign, the negative contribution to the $a_\mu$.
Thus, the $Z'$ should have either only a vector coupling or a dominant vector coupling plus a small axial coupling.

There are some dark force models corresponding to each case.
One is the dark photon model \cite{ArkaniHamed:2008qn}, which has only the vector coupling at the leading order.
Another is a relatively new model, the dark $Z$ model \cite{Davoudiasl:2012ag}, which has the vector coupling and a small axial coupling.
Their interactions are given by
\begin{eqnarray}
{\cal L}_{{\rm dark}~\gamma} &=& - \eps e J^\mu_{EM} Z'_\mu \label{eq:darkPhoton} \\
{\cal L}_{{\rm dark}~Z} &=& - \left[ \eps e J^\mu_{EM} + \eps_Z (g / 2 \cos\theta_W) J^\mu_{NC} \right] Z'_\mu \label{eq:darkZ}
\end{eqnarray}
with $J_\mu^{EM} = Q_f \bar f \gamma_\mu f$ and $J_\mu^{NC} = (T_{3f} - 2 Q_f \sin^2\theta_W ) \bar f \gamma_\mu f - (T_{3f}) \bar f \gamma_\mu \gamma_5 f$.
$\eps$ and $\eps_Z$ are the parametrizations of the effective $\gamma - Z'$ mixing and $Z - Z'$ mixing, respectively.

Both models commonly assume the kinetic mixing between the $U(1)_Y$ and the $U(1)_{\rm dark}$ \cite{Holdom:1985ag}.
The SM particles have zero charges under the new gauge group $U(1)_{\rm dark}$.
The gauge boson $Z'$ of the $U(1)_{\rm dark}$ can still couple to the SM fermions through the mixing with the SM gauge bosons.
Depending on the details of the Higgs sector, the physical eigenstate of the $Z'$ can mix only with the photon (dark photon model) or with both the photon and the $Z$ boson (dark $Z$ model).
Because of the $Z$ coupling, the $Z'$ in the dark $Z$ model inherits some properties of the $Z$ boson such as the parity violating nature.
To emphasize this $Z$ coupling, the $Z'$ of this model was named the {\em dark $Z$ boson}.
Roughly speaking, the dark photon is a heavier version of the photon, and the dark $Z$ is a lighter version of the $Z$ boson.

\begin{figure}[tb]
\begin{center}
\includegraphics[width=28pc]{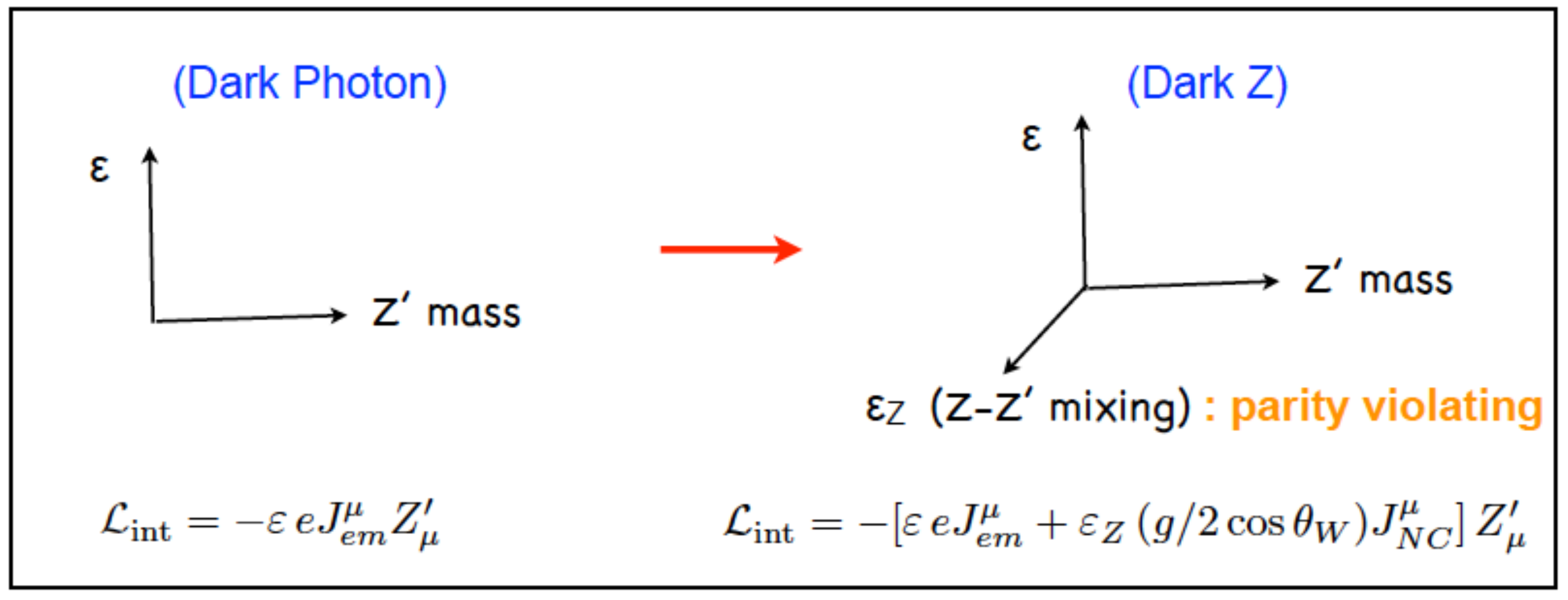}
\caption{\label{fig:DarkPhotonDarkZ}
Dark Photon vs. Dark $Z$. The dark $Z$ is a dark photon with more general couplings.}
\end{center}
\end{figure}

One of the effects of the dark $Z$ over the dark photon is that the relevant parameter space for the $Z'$ is extended from 2D to 3D (see the illustration in Fig.~\ref{fig:DarkPhotonDarkZ}).
In the dark photon model, we need only 2 parameters ($m_{Z'}$ and $\eps$), and all relevant couplings of the $Z'$ are determined by the $\eps$ [see Eq.~(\ref{eq:darkPhoton})].
In the dark $Z$ model, we have another parameter ($\eps_Z$), which describes the $Z - Z'$ mixing.
The coupling is given by the combination of the $\eps$ and $\eps_Z$ [see Eq.~(\ref{eq:darkZ})].
Thus, one can view the dark $Z$ as a dark photon with more general couplings, and the dark photon as a special case of the dark $Z$ in the $\eps_Z \to 0$ limit.

Because of the new coupling, some experiments that are irrelevant to the dark photon searches are relevant to the dark $Z$ searches \cite{Davoudiasl:2012ag,Davoudiasl:2012qa,Davoudiasl:2014kua,Kong:2014jwa,Davoudiasl:2013aya,Davoudiasl:2014mqa}.
They include the low-energy parity test, which we will discuss later in this paper.

\section{\boldmath Dark Force searches}
There are many ongoing and proposed searches for the dark force in the labs around the world \cite{Essig:2013lka}.
The typical searches exploit the small $Z'$ coupling to the SM particles rather than the coupling to the dark matter particles.
A particularly attractive feature about the dark force is that it is one of the rare new physics scenarios that can be tested/discovered at the low-energy experimental facilities, which are typically built for nuclear or hadronic physics.
It is possible because the dark force carrier $Z'$ is roughly of GeV scale, which is about 1000 times smaller than the typical new physics scale (TeV scale).
If a new particle is of TeV scale, the LHC has probably the best chance to find it, but the dark force carrier is of about proton mass scale, and it can be searched for in many low-energy labs.

\begin{figure*}[tb]
\begin{center}
\subfigure[]{
      \includegraphics[width=15pc,clip]{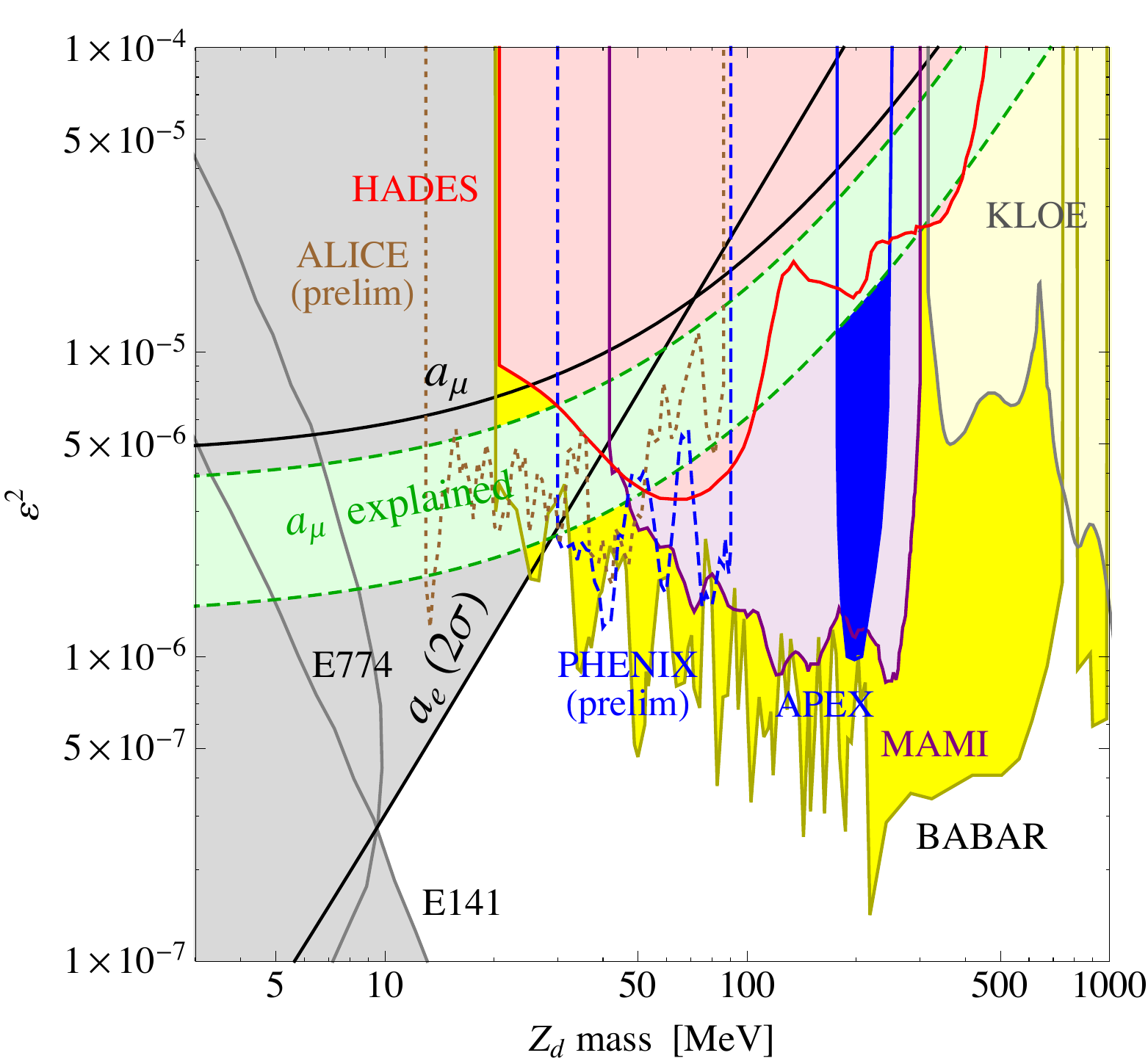}
      \label{fig:darkPhoton}
} \quad
\subfigure[]{
      \includegraphics[width=15pc,clip]{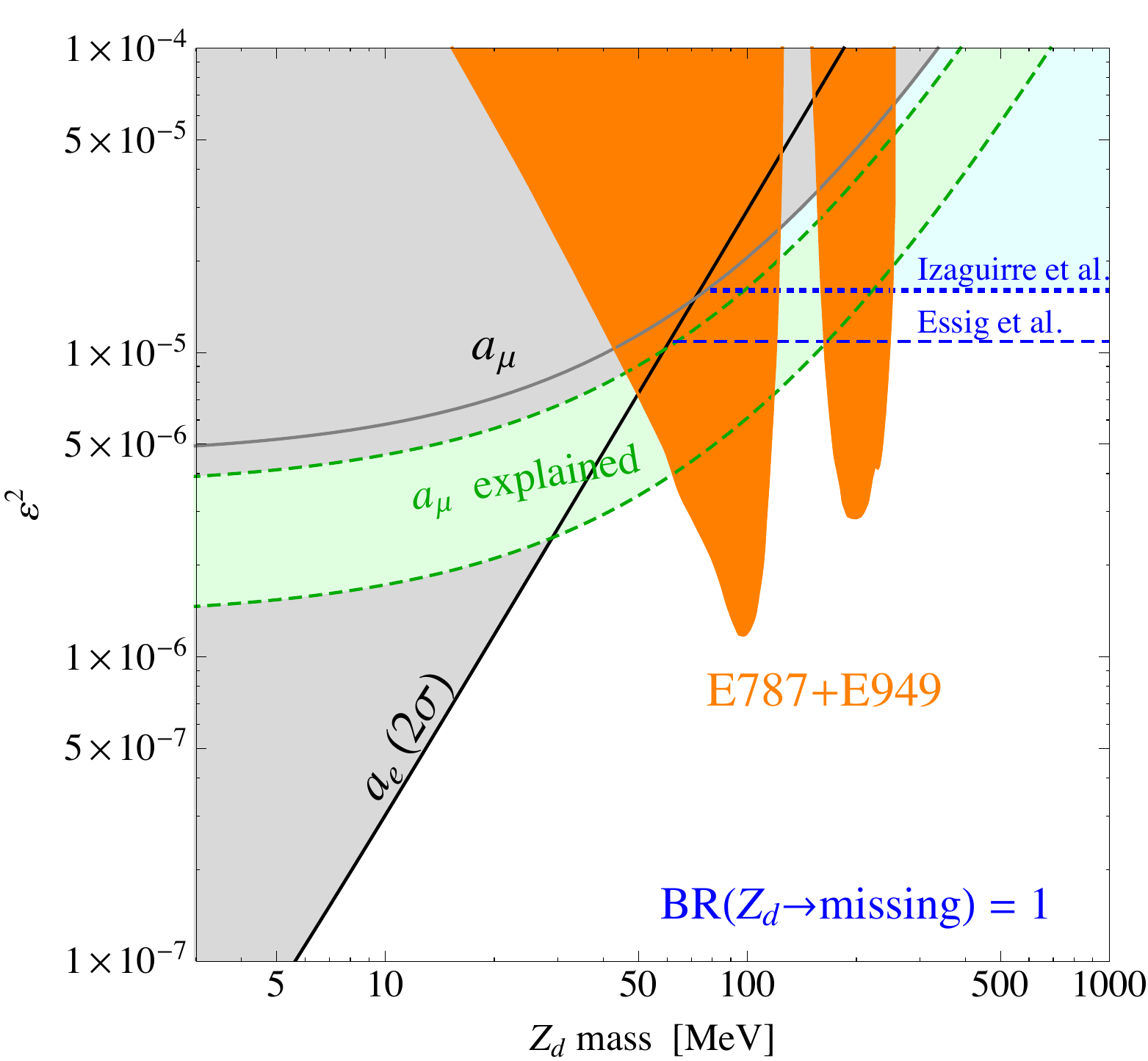}
      \label{fig:darkPhotonInvisible}
}
\caption{Dark photon parameter space in (a) the visible case and (b) the invisible case.}
\end{center}
\end{figure*}

Figure~\ref{fig:darkPhoton} (from Refs.~\cite{Davoudiasl:2014kua,Lee:2014tba} with a slight modification) shows the parameter space of the dark photon with the current bounds.
The bounds come from the electron \cite{Davoudiasl:2012ig,Endo:2012hp} and muon \cite{Gninenko:2001hx,Fayet:2007ua,Pospelov:2008zw} anomalous magnetic moments, fixed target experiments \cite{Abrahamyan:2011gv,Merkel:2014avp}, beam dump experiments \cite{Andreas:2012mt}, meson decays \cite{Bjorken:2009mm,Agakishiev:2013fwl,Babusci:2014sta,Lees:2014xha,Adare:2014mgk,ALICE}, $e^+e^-$ collision experiments \cite{Lees:2014xha}.
It is quite clear that with the most recent experimental dilepton bump searches along with the $a_e$ ($2 \sigma$ C.L.), the whole $g_\mu - 2$ favored parameter region (green band) is practically excluded.
Thus one of the leading motivations for the light dark gauge boson got seriously weakened.

Partly because of this reason, an invisibly-decaying dark photon has been spotlighted recently.
It assumes the existence of a very light dark matter ($\chi$) of $m_\chi < m_{Z'} / 2$ so that the $Z'$ can dominantly decay into the invisible particles.
Different experiments apply for this scenario.
Figure~\ref{fig:darkPhotonInvisible} (from Ref.~\cite{Davoudiasl:2014kua}) shows the parameter space of the invisibly-decaying dark photon with the current bounds.
It is basically the missing energy search.
The AGS $K \to \pi$ + nothing experiments \cite{Artamonov:2009sz} and the BABAR $e^+ e^- \to \gamma$ + nothing experiments \cite{Izaguirre:2013uxa,Essig:2013vha} constrain the parameter space.
There may be more constraints from the light dark matter scattering at detectors in the beam-dump experiments \cite{Diamond:2013oda,Batell:2014mga}, but they depend on the dark matter coupling, which is a unknown parameter.
We can see some of the parameter region ($m_{Z'} \sim 30 - 50 ~\mev$ and around $m_\pi$) survive the constraints.
These remaining parameter space may be explored by some proposed missing energy search experiments such as the DarkLight \cite{Kahn:2012br} and SPS beam-dump experiment \cite{Gninenko:2013rka,Andreas:2013lya}.

If we take the dark $Z$ model, things are a little different.
A visibly-decaying dark $Z$ has similar bounds as the dark photon in the shown parameter ranges [Fig.~\ref{fig:darkPhoton}].
For an invisibly-decaying dark $Z$, in the presence of the light dark matter particles, because of an additional term for the $Z - Z'$ mass mixing, there can be a sizable interference between the kinetic mixing ($\eps$) contribution and the mass mixing ($\eps_Z$) contribution in the flavor-changing meson decays.
The $K \to \pi$ + nothing constraints can be much weaker than that of the dark photon case [see Fig.~\ref{fig:darkZInvisible} from Ref.~\cite{Davoudiasl:2014kua}].

\section{\boldmath Dark Force search through Parity Violation Tests}
Now, we discuss the dark force searches through the low-energy parity test \cite{Davoudiasl:2012ag,Davoudiasl:2012qa,Davoudiasl:2014kua}.
The dark $Z$ modifies the effective lagrangian of the weak neutral current scattering,
\bea
{\cal L}_{\rm eff} &=& - \frac{4 G_F}{\sqrt{2}} J^\mu_{NC} ( \sin^2\theta_W ) J_\mu^{NC} ( \sin^2\theta_W ) \\
G_F &\to& \left(1 + \delta^2 \frac{1}{1 + Q^2 / m_{Z'}^2} \right) G_F \\
\sin^2\theta_W &\to& \left(1 - \eps\delta \frac{m_Z}{m_{Z'}} \frac{\cos\theta_W}{\sin\theta_W} \frac{1}{1 + Q^2 / m_{Z'}^2} \right) \sin^2\theta_W
\eea
where $Q$ is the momentum transfer of the two neutral currents, and $\delta$ is a reparametrization of the $\eps_Z$ with $\eps_Z \equiv (m_{Z'} / m_Z) \delta$.
One of the things these formulas indicate is these shifts are sensitive only to the low-$Q^2$ (low momentum transfer).
Thus, the dark $Z$ effectively changes the weak neutral current scattering, including the effective Weinberg angle, which describes the parity violation, but only for the low momentum transfer.

\begin{figure*}[tb]
\begin{center}
\subfigure[]{
      \includegraphics[width=15pc,clip]{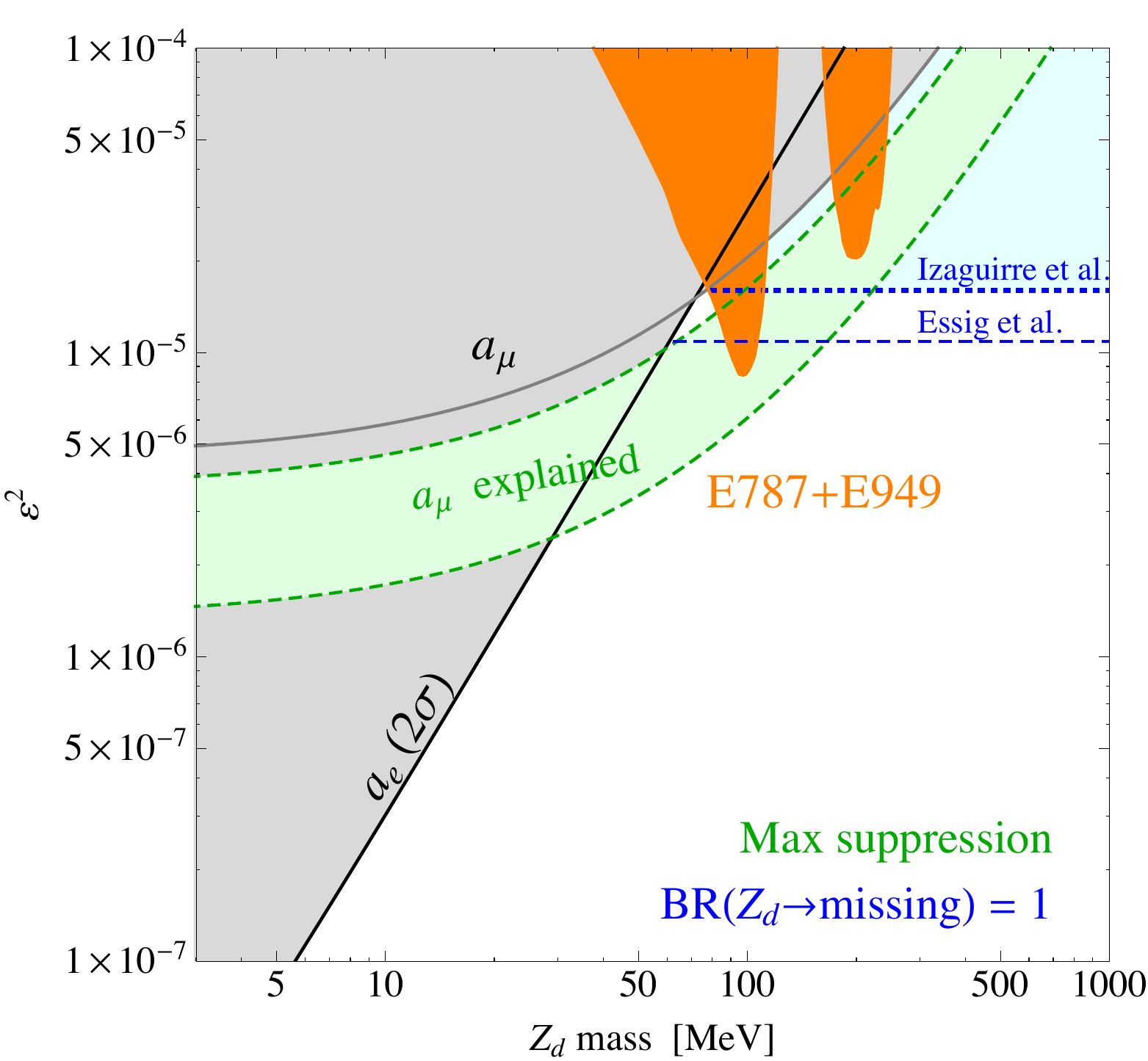}
      \label{fig:darkZInvisible}
} \quad
\subfigure[]{
      \includegraphics[width=20pc,clip]{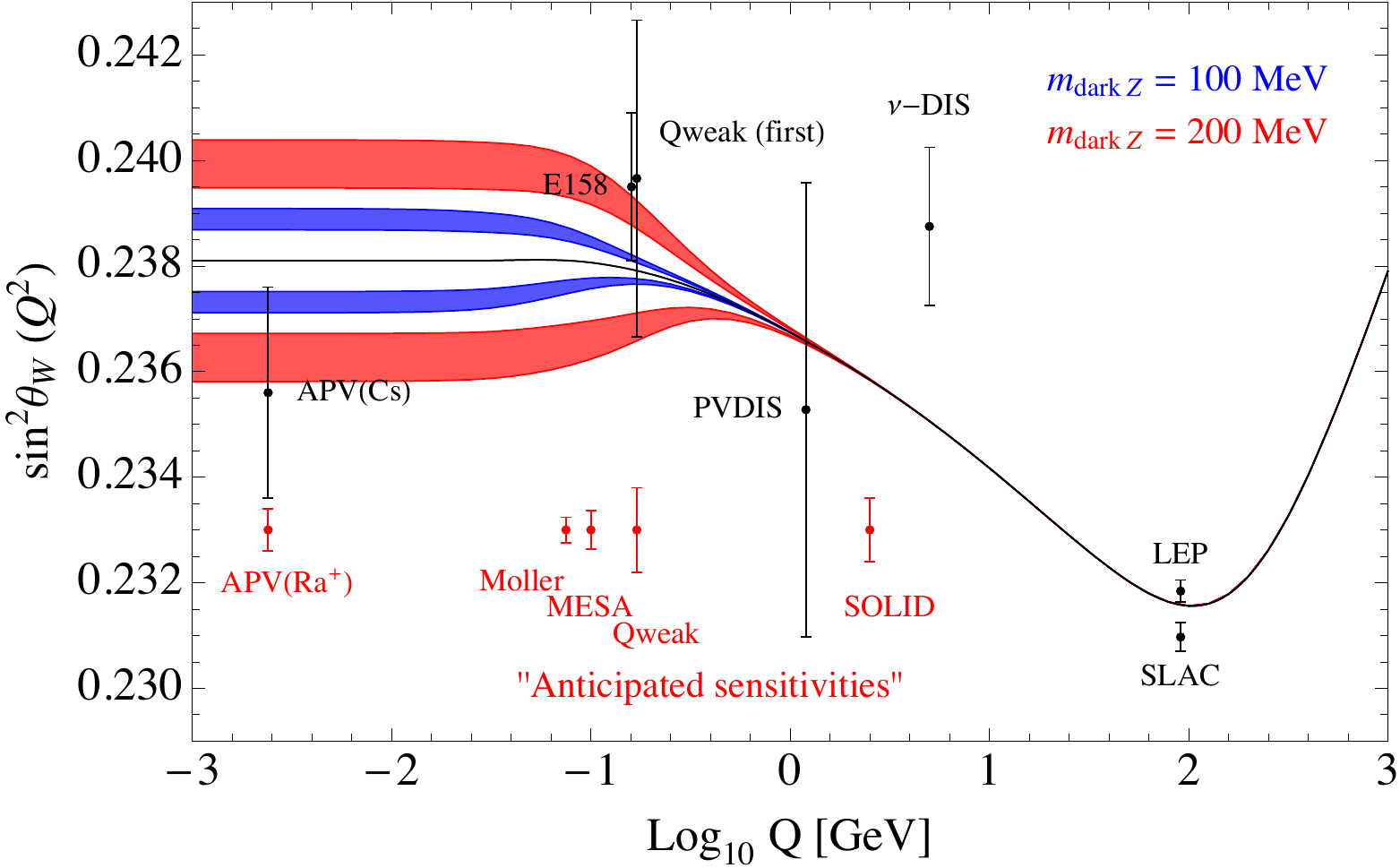}
      \label{fig:runningWeinberg}
}
\caption{(a) Parameter space of the invisibly-decaying dark $Z$. (b) Running of the effective $\sin^2\theta_W$ with $Q$. The black curve is the SM prediction, and the colored bands are new predictions.}
\end{center}
\end{figure*}

Figure~\ref{fig:runningWeinberg} (from Ref.~\cite{Davoudiasl:2014kua} with a slight modification) shows an example of how the effective $\sin^2\theta_W$ changes with $Q$ in the presence of a dark $Z$ for $m_{Z'} = 100~\mev$ (blue band), $200 ~\mev$ (red band) cases.
This example is actually for the invisibly-decaying dark $Z$.
Although the details are not very important for our discussions, these masses are those excluded in the (invisibly-decaying) dark photon case as a solution to the $g_\mu - 2$ anomaly, but in the dark $Z$, they are saved because of an additional $\eps_Z$ term, which predicts the specific parameter values giving the colored bands.
The deviations appear only in the low $Q$ values, roughly $Q \leq m_{Z'}$.
They never appears in the high $Q$ values relevant to the high-energy experiments.
In other words, we need low-energy experiments to see the dark $Z$ mediated scattering effects.

For the low-$Q^2$ parity tests, one can use the atomic parity violation in {\rm Cs} \cite{Gilbert:1986ki,Wood:1997zq,Bennett:1999pd}, {\rm Ra$^+$} ion \cite {NunezPortela2014,Jungmann:2014kia} or the low-$Q^2$ polarized electron scattering experiments E158 \cite{Anthony:2005pm}, Qweak \cite{Androic:2013rhu}, Moller \cite{Mammei:2012ph} and MESA \cite{MESA}.
The possible deviations due to the dark $Z$ can be large enough to be observed with the future experiments.
Deep inelastic scattering experiments such as PVDIS \cite{Wang:2014bba} and SOLID \cite{Reimer:2012uj} may be also useful for the heavier $Z'$.
The parity tests, measuring the Weinberg angle, are independent of the $Z'$ decay branching ratios, which makes them an excellent probe of the dark force for both visibly-decaying and invisibly-decaying $Z'$.

For some works on the low-$Q^2$ parity violations induced by a pseudoscalar particle, see Refs.~\cite{Stadnik:2013raa,Roberts:2014dda}.

\section{\boldmath Summary}
The dark gauge interaction has been being searched for in various ways.
(i) The $g-2$ of the electron and muon.
Especially, the $3.6 \sigma$ C.L. deviation in the $g_\mu - 2$ can be explained by the dark gauge boson (the green band in the parameter space).
(ii) The dilepton resonance search is probably the most popular and direct search of the dark gauge interaction, and it practically excludes the green band with the most recent data.
(iii) The missing energy search requires very light dark sector particles, but some parts of the green band survive.
(iv) The low-energy parity test, including the atomic parity violation and the polarized electron scatterings are simply another excellent search of the dark force, and it is independent of the $Z'$ decay branching ratios. It requires the presence of an axial coupling of the $Z'$ though as in the dark $Z$ model.

\ack{
We thank the PAVI 14 organizers for the invitation and hospitality.
We thank H. Davoudiasl and W. Marciano for the opportunity of collaboration in Dark Parity Violation, which this presentation is based on.
This work was supported in part by the CERN-Korea fellowship.
}

\section*{References}


\end{document}